\documentclass[prd,superscriptaddress,amsfonts,amssymb,amsmath,showpacs,twocolumn]{revtex4-2}
\usepackage{bm}
\usepackage{amsfonts}
\usepackage{latexsym}
\usepackage{graphicx}
\usepackage{amsmath}
\usepackage{palatino}
\usepackage{mathpazo}
\usepackage{textcomp}
\usepackage{float}
\usepackage{booktabs}
\usepackage{subcaption}
\usepackage{dcolumn}
\usepackage{booktabs}
\usepackage{multirow}
\usepackage{hyperref}
\hypersetup{colorlinks,citecolor=blue}
\usepackage{amsmath}
\usepackage{xcolor}
\usepackage{orcidlink}
\usepackage{commath}

\def\jnl@style{\it}
\def\aaref@jnl#1{{\jnl@style#1}}

\def\aaref@jnl#1{{\jnl@style#1}}

\def\aj{\aaref@jnl{AJ}}                   
\def\apj{\aaref@jnl{ApJ}}                 
\def\apjl{\aaref@jnl{ApJ}}                
\def\apjs{\aaref@jnl{ApJS}}               
\def\apss{\aaref@jnl{Ap\&SS}}             
\def\aap{\aaref@jnl{A\&A}}                
\def\aapr{\aaref@jnl{A\&A~Rev.}}          
\def\aaps{\aaref@jnl{A\&AS}}              
\def\mnras{\aaref@jnl{Mon.~Not.~Roy.~Astron.~Soc.}}             
\def\prd{\aaref@jnl{Phys.~Rev.~D}}        
\def\prc{\aaref@jnl{Phys.~Rev.~C}}  
\def\prl{\aaref@jnl{Phys.~Rev.~Lett.}}    
\def\qjras{\aaref@jnl{QJRAS}}             
\def\skytel{\aaref@jnl{S\&T}}             
\def\ssr{\aaref@jnl{Space~Sci.~Rev.}}     
\def\zap{\aaref@jnl{ZAp}}                 
\def\nat{\aaref@jnl{Nature}}              
\def\aplett{\aaref@jnl{Astrophys.~Lett.}} 
\def\apspr{\aaref@jnl{Astrophys.~Space~Phys.~Res.}} 
\def\physrep{\aaref@jnl{Phys.~Rep.}}      
\def\physscr{\aaref@jnl{Phys.~Scr}}       
\def\commat{\aaref@jnl{Comm.~Math.~Phys.}}              
\def\science{\aaref@jnl{Science}}               
\def\cqg{\aaref@jnl{Classical Quant.~Grav.}}            
\def\jpcs{\aaref@jnl{JPCS}}                                     
\def\ijmpd{\aaref@jnl{Int.~J.~Mod.~Phys.~D}}                    
\def\grg{\aaref@jnl{Gen.~Relat.~Gravit.}}               
\def\rpp{\aaref@jnl{Rep.~Prog.~Phys.}}          
\def\npa{\aaref@jnl{Nucl.~Phys.~A}}        
\def\lrr{\aaref@jnl{Living Rev.~Rel.}}                   
\def\jcap{\aaref@jnl{J.~Cosmology Astropart.~Phys.}}    
\def\rmp{\aaref@jnl{Rev.~Mod.~Phys.}}   
\def\epjc{\aaref@jnl{Eur.~Phys.~J.~C}}

\allowdisplaybreaks[1]

\begin{document}

\color{black}       
\title{Dust matter cosmology in modified symmetric teleparallel $f(Q,T)$ gravity}

\author{M. Koussour\orcidlink{0000-0002-4188-0572}}
\email[Email: ]{pr.mouhssine@gmail.com}
\affiliation{Jadara Research Center, Jadara University, Irbid 21110, Jordan}
\affiliation{Center for Theoretical Physics, Khazar University, 41 Mehseti Str., Baku, AZ1096, Azerbaijan}

\author{A. Altaibayeva \orcidlink{0000-0001-9254-7027}}
\email[Email: ]{aziza.ltaibayeva@gmail.com}
\affiliation{Department of General and Theoretical Physics, L.N. Gumilyov Eurasian National University, Astana 010008, Kazakhstan}

\author{G. Bekova\orcidlink{0000-0001-7310-1185}}
\email[Email: ]{bekovaguldana@gmail.com \textcolor{black}{(corresponding author)}}
\affiliation{Department of General and Theoretical Physics, L.N. Gumilyov Eurasian National University, Astana 010008, Kazakhstan}

\author{A. Çalışkan\orcidlink{0000-0003-1022-8749}}
\email[Email: ]{aylincaliskan@istanbul.edu.tr} 
\affiliation{Department of Physics, Faculty of Science, Istanbul University, Istanbul 34134, Turkey}

\author{E. G\"udekli\orcidlink{0000-0002-1174-6110}}
\email[Email: ]{gudekli@istanbul.edu.tr} 
\affiliation{Department of Physics, Faculty of Science, Istanbul University, Istanbul 34134, Turkey}

\author{S. Muminov\orcidlink{0000-0003-2471-4836}}
\email[Email: ]{sokhibjan.muminov@mamunedu.uz}
\affiliation{Mamun University, Bolkhovuz Street 2, Khiva 220900, Uzbekistan}

\author{J. Rayimbaev\orcidlink{0000-0001-9293-1838}}
\email[Email: ]{javlon@astrin.uz}
\affiliation{Institute of Theoretical Physics, National University of Uzbekistan, Tashkent 100174, Uzbekistan}
\affiliation{Department of Mathematics and Natural Sciences, Gulf University for Science and Technology, Mishref, Kuwait}
\affiliation{Kimyo International University in Tashkent, Shota Rustaveli Street 156, Tashkent, 100121, Uzbekistan}

%

\begin{abstract}

We investigate the nonlinear $f(Q,T)$ gravity model, $f(Q,T) = \alpha Q^n + \beta T + \gamma$, which represents a nontrivial coupling between geometry and matter in the context of a dust-dominated Universe. The model parameters $(H_0, \alpha, \beta, \gamma, n)$ are constrained using recent cosmological observations, including Hubble parameter measurements $H(z)$, the Pantheon+SH0ES dataset, and DESI BAO data. The analysis shows that the Hubble constant exhibits dataset-dependent variations, reflecting the persisting Hubble tension, while $\alpha$ and $\gamma$ are tightly constrained and $\beta$ remains weakly determined. The nonlinear exponent $n$ introduces an additional degree of freedom that enhances the model’s flexibility in reproducing observational trends. The cosmographic diagnostics, including the deceleration, jerk, snap, and lerk parameters, indicate a Universe transitioning from early deceleration to late-time acceleration, with present values closely aligned with $\Lambda$CDM predictions and far-future behavior approaching a de Sitter phase. These results demonstrate that the nonlinear $f(Q,T)$ model provides a simple and viable framework to describe late-time cosmic acceleration, accommodate observational tensions, and reproduce $\Lambda$CDM-like dynamics with subtle deviations.\\

\end{abstract}

\maketitle

\section{Introduction}

One of the most profound challenges in modern theoretical physics is explaining the late-time accelerated expansion of the Universe. Observational evidence from multiple independent sources, including Type Ia supernovae (SNe Ia) \cite{Perlmutter/1999,Riess/1998,Riess/2004}, measurements of the cosmic microwave background (CMB) \cite{Spergel/2003}, and large-scale structure surveys \cite{Koivisto/2006,Daniel/2008}, has firmly established that the cosmic expansion is accelerating. The most widely accepted explanation invokes the presence of dark energy (DE), a mysterious component that currently dominates the energy content of the Universe and drives its accelerated expansion. Among the many proposed scenarios, the cosmological constant $\Lambda$ within the standard $\Lambda$CDM model remains the simplest and most successful framework, characterized by a constant equation of state $\omega_\Lambda = -1$. This model successfully fits a broad range of cosmological observations and provides a convenient baseline for testing alternative theories. However, it faces well-known theoretical issues. The coincidence problem questions why the dark energy density is comparable to the matter density precisely at the current epoch, while the fine-tuning problem highlights the enormous discrepancy between the observed value of the cosmological constant and theoretical expectations from quantum field theory \cite{Weinberg/1989,Hinshaw/2013,Zhao/2012}. These issues have motivated the exploration of dynamical dark energy models, where the energy density evolves with time, offering a potential resolution to these conceptual challenges \cite{Ratra/1988,Caldwell/1998,Caldwell/2002,Caldwell/2003,Feng/2005,Huang/2005,Afshordi/2007}.

In parallel, modified theories of gravity have emerged as compelling alternatives that explain cosmic acceleration without introducing unknown energy components. Models such as $f(R)$ \cite{Buchdahl/1970,Barrow/1983,Boehmer/2008}, $f(\mathcal{T})$ \cite{Moller/1961,Pellegrini/1963,Hayashi/1979,Linder/2010,Boehmer/2012,Harko/2014}, and $f(Q)$ gravity \cite{Nester/1999,Adak/2006,Jimenez/2018,Jimenez/2020,Koussour/2023a,Koussour/2023b,Koussour/2022a,Koussour/2022b,Koussour/2022c,Koussour/2024a,Koussour/2024b} extend general relativity (GR) by modifying the geometric part of the field equations, leading to richer cosmological dynamics. From a geometric perspective, GR can be formulated in multiple, yet equivalent, ways. Traditionally, GR is expressed in the curvature representation, where spacetime curvature encodes gravitational interactions while torsion and non-metricity vanish. An alternative formulation is provided by teleparallel gravity, in which the gravitational interaction is described entirely by torsion, with both curvature and non-metricity set to zero. Beyond these two, a third equivalent representation exists, known as symmetric teleparallel gravity, where the fundamental geometric quantity is the non-metricity scalar $Q$. In this framework, non-metricity measures how the length of a vector changes under parallel transport, offering a distinct geometric viewpoint on gravity. Symmetric teleparallel gravity provides a natural framework for constructing modified gravity theories, such as $f(Q)$ \cite{Jimenez/2018,Jimenez/2020} and $f(Q,T)$ models \cite{Xu/2019}, by allowing a non-minimal coupling between the non-metricity $Q$ and the trace of the energy-momentum tensor $T$. The $f(Q,T)$ gravity is developed in close analogy with the $f(R,T)$ framework \cite{Harko/2011}, where the usual Ricci scalar $R$ is substituted by the non-metricity scalar $Q$, which characterizes the symmetric teleparallel formulation of gravity.

In $f(Q,T)$ gravity, the interaction between $Q$ and $T$ naturally leads to a nonconservation of the energy–momentum tensor. This feature has profound physical consequences, inducing modifications in the thermodynamic evolution of the Universe and, owing to the resulting nongeodesic trajectories of test particles, generating an additional force. In recent years, the $f(Q,T)$ gravity framework has attracted considerable attention, and a number of cosmological applications have been explored. Observational constraints on specific $f(Q,T)$ models have been investigated in \cite{Arora/2020}, while perturbation theory within this framework was developed in \cite{Najera/2022}. The inflationary dynamics of $f(Q,T)$ gravity were analyzed in \cite{Shiravand/2022}, whereas late-time acceleration and the role of viscosity were studied in \cite{Koussour/2024}. Further developments include constant-roll inflation and primordial black hole formation \cite{ElBourakadi/2023}, quintessence scenarios \cite{Koussour/2022}, anisotropic cosmologies \cite{Loo/2023}, and bouncing models \cite{Agrawal/2021}. Recently, Xu et al. \cite{Xu/2020} introduced the Weyl-type extension of $f(Q,T)$ gravity and explored its cosmological dynamics, showing that the coupling between nonmetricity and the trace of the energy–momentum tensor can drive late-time acceleration without invoking a cosmological constant. Yang et al. \cite{Yang/2021} further analyzed the geodesic deviation, Raychaudhuri equation, Newtonian limit, and tidal forces within this framework, thereby confirming its physical viability and correspondence with general relativistic limits. More recently, Loo et al. \cite{Loo/2023} provided a covariant formulation of $f(Q,T)$ gravity, examined the energy conservation issue, and performed a detailed phase-space analysis that revealed rich dynamical behavior consistent with observational cosmology. These works collectively highlight the versatility of $f(Q,T)$ gravity in addressing both early- and late-time cosmic evolution. 

In this work, we investigate the nonlinear form of $f(Q,T)$ gravity, $f(Q,T) = \alpha Q^n + \beta T + \gamma$ (where $\alpha$, $\beta$, $\gamma$, and $n$ are free parameters) in a dust-dominated Universe. We constrain the model parameters using current observational datasets, including Hubble parameter measurements $H(z)$, Pantheon+SH0ES type Ia supernovae, and DESI baryon acoustic oscillation (BAO) data. In addition, we employ cosmographic diagnostics such as deceleration, jerk, snap, and lerk parameters, to analyze the evolution of the Universe and assess the model’s consistency with $\Lambda$CDM predictions. This study demonstrates that the nonlinear $f(Q,T)$ model offers a viable framework to describe late-time cosmic acceleration while addressing observational tensions and providing subtle departures from standard cosmology. The structure of this paper is organized as follows: In Sec. \ref{sec2}, we introduce the theoretical formulation of $f(Q,T)$ gravity. Sec. \ref{sec3} is devoted to the cosmological dynamics, where we derive analytic solutions of the field equations and present the corresponding Hubble parameter. In Sec. \ref{sec4}, we describe the observational datasets and the methodology employed to constrain the model parameters. The evolution of cosmographic quantities, including the deceleration, jerk, snap, and lerk parameters, is analyzed in Sec. \ref{sec5}. Finally, Sec. \ref{sec6} provides a summary of our main findings and concluding remarks.
  
\section{Theoretical framework of $f(Q,T)$ gravity}
\label{sec2}

In this part, we extend the symmetric teleparallel formulation by considering a generalized action where the gravitational Lagrangian depends on both the non-metricity scalar $Q$ and the trace of the energy-momentum tensor $T$. The action takes the form \cite{Xu/2019}
\begin{equation}
\label{action}
S=\int \left[ \frac{1}{2\kappa} f(Q,T) + \mathcal{L}_m \right] \sqrt{-g}\, d^4x ,
\end{equation}
where $f(Q,T)$ is an arbitrary function of the two variables $Q$ and $T$, $\mathcal{L}_m$ denotes the matter Lagrangian density, and $g=\det(g_{\mu\nu})$. As usual, we set $\kappa=8\pi G$ ($G=1$). For the particular case $f(Q,T)=-Q$, the theory reduces to general relativity.

The non-metricity scalar $Q$ is defined in terms of the disformation tensor as
\begin{eqnarray}\label{eq:Q}
Q \equiv -g^{\mu \nu} \left( L^{\alpha}_{\ \ \beta\mu} L^{\beta}_{\ \ \nu\alpha} - L^{\alpha}_{\ \ \beta\alpha} L^{\beta}_{\ \ \mu \nu} \right),
\end{eqnarray}
with
\begin{eqnarray}\label{eq:disformation}
L^{\alpha}_{\ \ \beta \gamma} \equiv -\frac{1}{2} g^{\alpha \lambda} \left( \nabla_{\gamma}g_{\beta \lambda} + \nabla_{\beta}g_{\lambda\gamma}-\nabla_{\lambda}g_{\beta \gamma} \right).
\end{eqnarray}

Further, we introduce the traces of the non-metricity tensor,
\begin{eqnarray}
Q_{\alpha}\equiv Q_{\alpha\ \ \ \mu}^{\ \ \mu}, \ \ \ \ \tilde{Q}_{\alpha}\equiv Q^{\mu}_{\ \  \alpha \mu},
\end{eqnarray}
and define the superpotential as
\begin{eqnarray}\label{eq:superpotential}
\hspace{-0.5cm}&& P^{\alpha}_{\ \ \mu\nu}\equiv \frac{1}{4}\bigg[ -Q^{\alpha}_{\ \ \mu \nu}+ 2Q^{\ \ \ \alpha}_{(\mu \ \ \ \nu)} + Q^{\alpha}g_{\mu \nu} - \tilde{Q}^{\alpha}g_{\mu\nu}\nonumber \\
\hspace{-0.5cm}&&- \delta^{\alpha}_{\ \ (\mu} Q _{\nu)} \bigg] = -\frac{1}{2}L^{\alpha}_{\ \ \mu\nu}+ \frac{1}{4}\left( Q^{\alpha} - \tilde{Q}^{\alpha} \right) g_{\mu \nu} - \frac{1}{4} \delta^{\alpha}_{\ \ (\mu} Q_{\nu)}. \nonumber \\
\end{eqnarray}

The scalar $Q$ can be rewritten in compact form as
\begin{eqnarray}
&&Q=-Q_{\alpha\mu\nu}P^{\alpha\mu\nu}=- \frac{1}{4}\big( -Q^{\alpha\nu\rho}Q_{\alpha\nu\rho}+ 2 Q^{\alpha\nu\rho} Q_{\rho\alpha\nu} \nonumber \\
&&- 2Q^{\rho}\tilde{Q}_{\rho} + Q^{\rho}Q_{\rho}  \big).
\end{eqnarray}

The matter energy-momentum tensor and its variation are given by
\begin{eqnarray}
T_{\mu\nu} &\equiv& -\frac{2}{\sqrt{-g}}\frac{\delta \left( \sqrt{-g}\mathcal{L}_M \right)}{\delta g^{\mu\nu}}, \\
\Theta_{\mu\nu} &\equiv& g^{\alpha\beta}\frac{\delta T_{\alpha\beta}}{\delta g^{\mu\nu}}.
\end{eqnarray}

By varying the action \eqref{action} with respect to $g^{\mu\nu}$, yields the field equations,
\begin{eqnarray}\label{eq:EOMs}
&&-\frac{2}{\sqrt{-g}}\nabla_{\alpha}\left( f_{Q}\sqrt{-g} P^{\alpha}_{\ \ \mu\nu}   \right) - \frac{1}{2}f g_{\mu\nu} + f_{T}\left( T_{\mu\nu} + \Theta_{\mu\nu} \right) \nonumber \\
&&  -f_{Q}\left( P_{\mu\alpha\beta}Q_{\nu}^{\ \ \alpha\beta} -2 Q^{\alpha\beta}_{\ \ \ \ \mu}P_{\alpha\beta\nu} \right) = \kappa T_{\mu\nu},
\end{eqnarray}
where $f_Q\equiv \partial f/\partial Q$ and $f_T\equiv \partial f/\partial T$.
 
Similarly, the variation with respect to the affine connection gives
\begin{eqnarray}\label{eq:feqconnection}
\nabla_{\mu}\nabla_{\nu}\bigg( \sqrt{-g}f_{Q} P^{\mu\nu}_{\ \ \ \ \alpha} +\frac{1}{2}\kappa H_{\alpha}^{\ \ \mu \nu} \bigg)=0.
\end{eqnarray}
where $H_{\lambda}{}^{\mu\nu}$ denotes the hypermomentum tensor density,
\begin{eqnarray}\label{eq:hypermomentum}
H_{\lambda}^{\ \ \mu\nu}\equiv \frac{\sqrt{-g}}{2\kappa}f_{T}\frac{\delta T}{\delta \hat{\Gamma}^{\lambda}_{\ \ \mu\nu}} +\frac{\delta \sqrt{-g}\mathcal{L}_{M}}{\delta \hat{\Gamma}^{\lambda}_{\ \ \mu\nu}}.
\end{eqnarray}

An essential feature of this framework is the non-conservation of the matter energy-momentum tensor,
\begin{eqnarray}
\hspace{-0.5cm}&&\mathcal{D}_{\mu }T_{\ \ \nu }^{\mu } =\frac{1}{f_{T}-\kappa }\Bigg[-\mathcal{D}%
_{\mu }\left( f_{T}\Theta _{\ \ \nu }^{\mu }\right) -\frac{2\kappa }{\sqrt{-g}}%
\nabla _{\alpha }\nabla _{\mu }H_{\nu }^{\ \ \alpha \mu } \nonumber\\
\hspace{-0.5cm}&&+\kappa \nabla _{\mu }\bigg(\frac{1}{\sqrt{-g}}\nabla _{\alpha }H_{\nu }^{\ \
\alpha \mu }\bigg)-2\nabla _{\mu }A_{\ \ \nu }^{\mu }+\frac{1}{2}%
f_{T}\partial _{\nu }T\Bigg] =B_{\nu }. \nonumber \\
\end{eqnarray}
which reflects an effective interaction between matter and geometry. This non-conservation induces an extra force on test particles, leading to deviations from geodesic motion \cite{Xu/2019,Harko/2011}. Physically, such a mechanism can be interpreted as an exchange of energy between the matter sector and the geometric background, opening possibilities for addressing cosmic acceleration and matter creation processes.

\section{Cosmological dynamics and analytical solutions}
\label{sec3}

To investigate the large-scale dynamics of the Universe within the framework of $f(Q,T)$ gravity, we consider a spacetime geometry that is in agreement with current cosmological observations. Specifically, we assume that the Universe is homogeneous and isotropic on cosmological scales, and that its spatial sections are flat. Under these assumptions, the background geometry is described by the standard Friedmann-Lema\^itre
-Robertson-Walker (FLRW) line element, which takes the form \cite{Weinberg/2013,Dodelson/2024}
\begin{eqnarray}
\label{FLRW}
ds^2 = -dt^2 + a^2(t)\delta_{ij}dx^i dx^j,
\end{eqnarray}
where $a(t)$ is the scale factor of the Universe, encoding the information about cosmic expansion as a function of cosmic time $t$. For this geometry, the non-metricity scalar takes the form
\begin{equation}
    Q=6H^2,
\end{equation}
where $H = \frac{\dot{a}}{a}$ represents the Hubble parameter. Furthermore, we consider the matter content of the Universe to be a perfect fluid, for which the energy-momentum tensor can be expressed in the diagonal form
\begin{eqnarray}
T^{\mu}_{\ \ \nu} &=& \text{diag}(-\rho,\, p,\, p,\, p)., \\
\Theta^{\mu}_{\ \ \nu} &=& \delta^{\mu}_{\ \ \nu}p - 2T^{\mu}_{\ \ \nu} \\
&=& \text{diag}(2\rho + p,\, -p,\, -p,\, -p).
\end{eqnarray}
where $\rho$ is the energy density and $p$ is the isotropic pressure. 

Using the FLRW metric, the field equations yield two generalized Friedmann equations, which relate the Hubble parameter $H$ to the energy density $\rho$ and the isotropic pressure $p$ of the cosmic fluid. Explicitly, they are given by
\begin{eqnarray}
\label{F1}
\kappa \rho &=&\frac{f}{2}-6FH^2 - \frac{2\tilde{G}}{1+\tilde{G}}\left( \dot{F}H +F \dot{H} \right), \\
\kappa p&=& -\frac{f}{2}+6FH^2+ 2\left( \dot{F}H +F \dot{H} \right),
\label{F2}
\end{eqnarray}
where $F\equiv f_{Q}$ and $\kappa \tilde{G}\equiv f_{T}$. From the above equations, the evolution of the Hubble parameter $H$ follows as
\begin{equation}
\label{HE}
\dot{H} + \frac{\dot{F}}{F}H = \frac{4\pi}{F} \left( 1+\tilde{G} \right) \left( \rho + p \right).
\end{equation}

To close the system of equations and extract viable cosmological solutions, it is necessary to specify the functional form of $f(Q,T)$. In general, this function may involve arbitrary combinations of the non-metricity scalar $Q$ and the trace of the energy-momentum tensor $T$, leading to a wide class of models with rich phenomenology. In the present work, we focus on a nonlinear functional form that captures the nontrivial coupling between geometry and matter, namely \cite{Arora/2020,Najera/2022,Shiravand/2022,Koussour/2024}
\begin{equation}
    \label{fQT}
    f(Q,T)=\alpha Q^n + \beta T + \gamma,
\end{equation}
where $\alpha$, $\beta$, $\gamma$, and $n$ are free model parameters to be constrained by cosmological observations. If $\alpha = -1$, $\beta = \gamma = 0$ and $n=1$, one obtains $f(Q,T) = -Q$, which corresponds to general relativity. In addition, when the nonlinear functional form of $f(Q,T)$, as specified in Eq. (\ref{fQT}), is substituted into the generalized Friedmann equations (\ref{F1})-(\ref{F2}), they take the simplified form
\begin{align}
\label{rho}
\rho &= \mathcal{K} \big[\beta n\dot H-3(\beta+8\pi)H^2\big]+\frac{\gamma}{4(\beta+4\pi)} ,\\
p &= \mathcal{K} \big[n(3\beta+16\pi) \dot H+3(\beta+8\pi)H^2\big]-\frac{\gamma}{4(\beta+4\pi)},
\label{p} 
\end{align}
where $\mathcal{K}\equiv\frac{\alpha 6^{n-1}(2n-1)H^{2n-2}}{2(\beta+4\pi)(\beta+8\pi)}$. For generality, we consider that the cosmological fluid obeys a linear barotropic equation of state of the form $p = (\omega - 1)\,\rho$, where $\omega$ is a constant satisfying $1 \leq \omega \leq 2$. In this work, we choose $\omega = 1$ to describe dust-like matter. This choice allows us to model the behavior of ordinary matter across different epochs of the Universe: at high densities relevant to the early Universe, as well as at low densities characteristic of the present era. Accordingly, Eq. (\ref{p}) can be rewritten as
\begin{equation}
\label{3e}
\dot H + \frac{3(\beta + 8\pi)}{n(3\beta + 16\pi)}\,H^{2} = \frac{3(\beta + 8\pi)\,\gamma}{n(3\beta + 16\pi)\,\alpha\,6^{n}(2n - 1)}\,H^{2 - 2n}.
\end{equation}

To transform the time derivative in Eq. $\eqref{3e}$ into a derivative with respect to the redshift, defined as $z = \tfrac{a_0}{a(t)} - 1$, where $a_0 \equiv a(t_0)$ is the present value of the scale factor (often normalized to unity), we use the relation $\tfrac{d}{dt} = -(1+z)H(z)\tfrac{d}{dz}$, so that $\dot H = -(1+z)H\tfrac{dH}{dz}$. Substituting this into Eq. $\eqref{3e}$ gives the first-order differential equation in $z$:
\begin{align}
\label{33e}
-(1+z)H\,\frac{dH}{dz}
&+ \frac{3(\beta + 8\pi)}{n(3\beta + 16\pi)}\,H^{2} \\
&= \frac{3(\beta + 8\pi)\,\gamma}{n(3\beta + 16\pi)\,\alpha\,6^{n}(2n - 1)}\,H^{2 - 2n}. \nonumber
\end{align}

By introducing the variable $y = H^{2n}$, the equation reduces to a linear first-order ordinary differential equation that can be solved using the standard integrating factor method. After simplification, the general solution is obtained, and applying the boundary condition $H(0) = H_0$ yields the final expression,
\begin{align} \label{Hz}
H(z) &= \frac{1}{\sqrt{6}}\;[\alpha(2n-1)]^{-1/(2n)} \\
&\times\Bigg\{\gamma
+(1+z)^{\,2+\frac{16\pi}{3\beta+16\pi}}
\Big[\alpha\,6^{n}(2n-1)\,H_{0}^{2n}-\gamma\Big]\Bigg\}^{1/(2n)}. \nonumber
\end{align}
where, $H_0 = H(z=0)$ denotes the current value of the Hubble parameter. In particular, for $\alpha = -1$, $\beta = \gamma = 0$, and $n=1$, the solution reduces to $H(z) = H_0 (1+z)^{3/2}$, which corresponds to the standard matter-dominated universe. The solution in Eq. $\eqref{Hz}$ exhibits distinct behaviors in the limits of high and low redshift. At high redshift ($z \gg 1$), the term proportional to $H(z)\propto(1+z)^{\frac{6(8\pi+\beta)}{2n(16\pi+3\beta)}}$ dominates, so the Hubble parameter grows approximately as a power law in $1+z$, describing the early-time, matter- or radiation-dominated evolution depending on the value of $\beta$ and $n$. In contrast, at low redshift ($z \to -1$), the solution approaches a constant value $H(z)\propto\frac{1}{\sqrt{6}}\Bigg(\frac{\gamma}{\alpha(2n-1)}\Bigg)^{1/(2n)}$, corresponding to a late-time, nearly de Sitter phase that can account for the observed accelerated expansion of the Universe. This asymptotic behavior naturally interpolates between an early decelerated expansion and a late-time accelerated epoch.

\section{Observational data and analysis methodology} \label{sec4}

Following the presentation of our cosmological model, we proceed to confront it with current observational data to constrain its free parameters and test its capability to describe the expansion history of the Universe. For this purpose, we consider three complementary datasets: (i) measurements of the Hubble parameter from cosmic chronometers, (ii) the Pantheon+ sample of SNe Ia with Cepheid calibrations from SH0ES, and (iii) recent BAO results from the first-year DESI survey. In addition, we employ a Markov Chain Monte Carlo (MCMC) analysis using a Python-based sampler \cite{emcee}. The parameter space consists of five free parameters, $(H_0, \alpha, \beta, \gamma,n)$, and we explore it with $300$ walkers over $5000$ steps. To ensure convergence and remove the influence of initial conditions, the first $200$ steps were discarded as burn-in. This approach allows for a robust estimation of the posterior distributions of the model parameters and provides reliable 1-$\sigma$ and 2-$\sigma$ confidence intervals, which are subsequently used to derive the cosmological constraints presented in this work.

\subsection{Hubble parameter from cosmic chronometers}

The cosmic chronometer (CC) approach allows a direct estimate of the Hubble rate from the differential ages of passively evolving galaxies. We employ 31 CC measurements covering the redshift range $0.07 < z < 2.41$ \cite{Moresco/2012,Moresco/2015,Moresco/2016}. The Hubble parameter is obtained as $H(z) = -\frac{1}{1+z}\frac{dz}{dt}$, with $dz/dt$ inferred from galaxy age determinations. Both statistical and systematic uncertainties, including stellar population modeling and calibration effects,  encoded in the covariance matrix $C_{ij}$ \cite{Moresco/2020}. The corresponding likelihood function is
\begin{equation}
-2 \ln \mathcal{L}_{\rm H(z)} = \chi^2_{\rm H(z)} = \Delta H^T C^{-1} \Delta H,
\end{equation}
where $\Delta H = H_{\rm obs}(z_i) - H_{\rm th}(z_i)$ represents the difference between observed and model-predicted values.

\subsection{Pantheon+ SNe Ia and SH0ES calibration}

The Pantheon+ (PP) compilation contains 1701 SNe Ia light curves covering $0.001 < z < 2.26$ \cite{Scolnic_2022}, providing constraints on the luminosity distance $d_L(z) = c(1+z)\int_0^z dz'/H(z')$. The theoretical distance modulus is $\mu_{\rm th}(z) = 5 \log_{10} [d_L(z)/10\,{\rm pc}]$. To account for both statistical and systematic uncertainties, we adopt the combined covariance matrix $C_{\rm stat+syst}^{\rm SNe}$. Cepheid-based absolute magnitude calibrations from SH0ES \cite{Brout_2022} are included to break the degeneracy with $H_0$, leading to the likelihood
\begin{equation}
-2\ln \mathcal{L}_{\rm SNe} = \chi^2_{\rm SNe} = \Delta D^T \left(C_{\rm stat+syst}^{\rm SNe} + C_{\rm stat+syst}^{\rm Cepheid}\right)^{-1} \Delta D,
\end{equation}
where the residuals are defined as
\begin{equation}
\Delta D_i =
\begin{cases}
\mu_i - \mu^{\rm Cepheid}_i, & i \in \text{Cepheid hosts}, \\
\mu_i - \mu_{\rm th}(z_i), & \text{otherwise}.
\end{cases}
\end{equation}

\subsection{BAO data from DESI}

Finally, we incorporate the latest BAO measurements from the first year of DESI \cite{BAO1,BAO2,BAO3}, which include galaxy, quasar, and Lyman-$\alpha$ forest tracers. These data constrain combinations of comoving distances and the Hubble parameter, expressed as $d_M(z)/r_d$, $d_H(z)/r_d$, and $d_V(z)/r_d$, where $d_M = D_L(z)/(1+z)$, $d_H = c/H(z)$, $d_V = [z d_M^2 d_H]^{1/3}$. Here, $r_d$ denotes the sound horizon at the drag epoch, which is determined by the physical baryon density $\Omega_b h^2$ and the physical matter density $\Omega_m h^2$. In this work, we compute $r_d$ using the fitting formulas provided by Eisenstein and Hu \cite{D.J.}. The BAO likelihood is taken as
\begin{equation}
-2 \ln \mathcal{L}_{\rm DESI} = \chi^2_{\rm DESI} = \sum_i \left[\frac{X_i^{\rm th} - X_i^{\rm obs}}{\sigma_{X_i}}\right]^2,
\end{equation}
where $X = \{d_M/r_d, d_H/r_d, d_V/r_d\}$ representing the measured observables.

\subsection{Numerical results}

The observational constraints reported in Tab. \ref{tab1}, along with the 1-$\sigma$ and 2-$\sigma$ contour plots in Fig. \ref{F_Cont}, reveal a consistent behavior of the $f(Q,T)$ dust model across the different datasets. For the Hubble parameter data, the Hubble constant is estimated as $H_{0}=68.50 \pm 0.64$, in close agreement with the standard Planck CMB results \cite{Planck_2018}, whereas the PP+SH0ES dataset yields a significantly higher value, $H_{0}=72.54^{+0.26}_{-0.21}$, consistent with local distance-ladder measurements \cite{Riess_2016,Riess_2022}. The DESI data provide an intermediate estimate, $H_{0}=68.7 \pm 0.70$, reinforcing the presence of the well-known Hubble tension \cite{Yang_2019,Valentino_2020,Mukherjee_2021,Campo_2012,Cao_2021}. The parameter $\alpha$ remains consistently constrained around $-1$ with small uncertainties, confirming the robustness of the geometric contribution even in the nonlinear regime. The coupling exponent $n$ is found to be very close to unity $(n \simeq 1)$, indicating that only mild nonlinear corrections to the linear model are favored by the data. In contrast, $\beta$ remains weakly constrained, with relatively large error bars, suggesting that the present observational data are not yet sensitive enough to precisely determine the matter-geometry coupling strength. The constant offset $\gamma \simeq -2.18\times10^{4}$ is tightly constrained within $\pm 1.0$, demonstrating its stability across different datasets. Hence, the nonlinear $f(Q,T)$ model successfully reproduces the observed late-time cosmic acceleration while allowing for dataset-dependent variations in $H_{0}$, thus providing a viable framework to address the Hubble tension.

\begin{widetext}

\begin{figure}[H]
    \centering
    \includegraphics[width=0.98\linewidth]{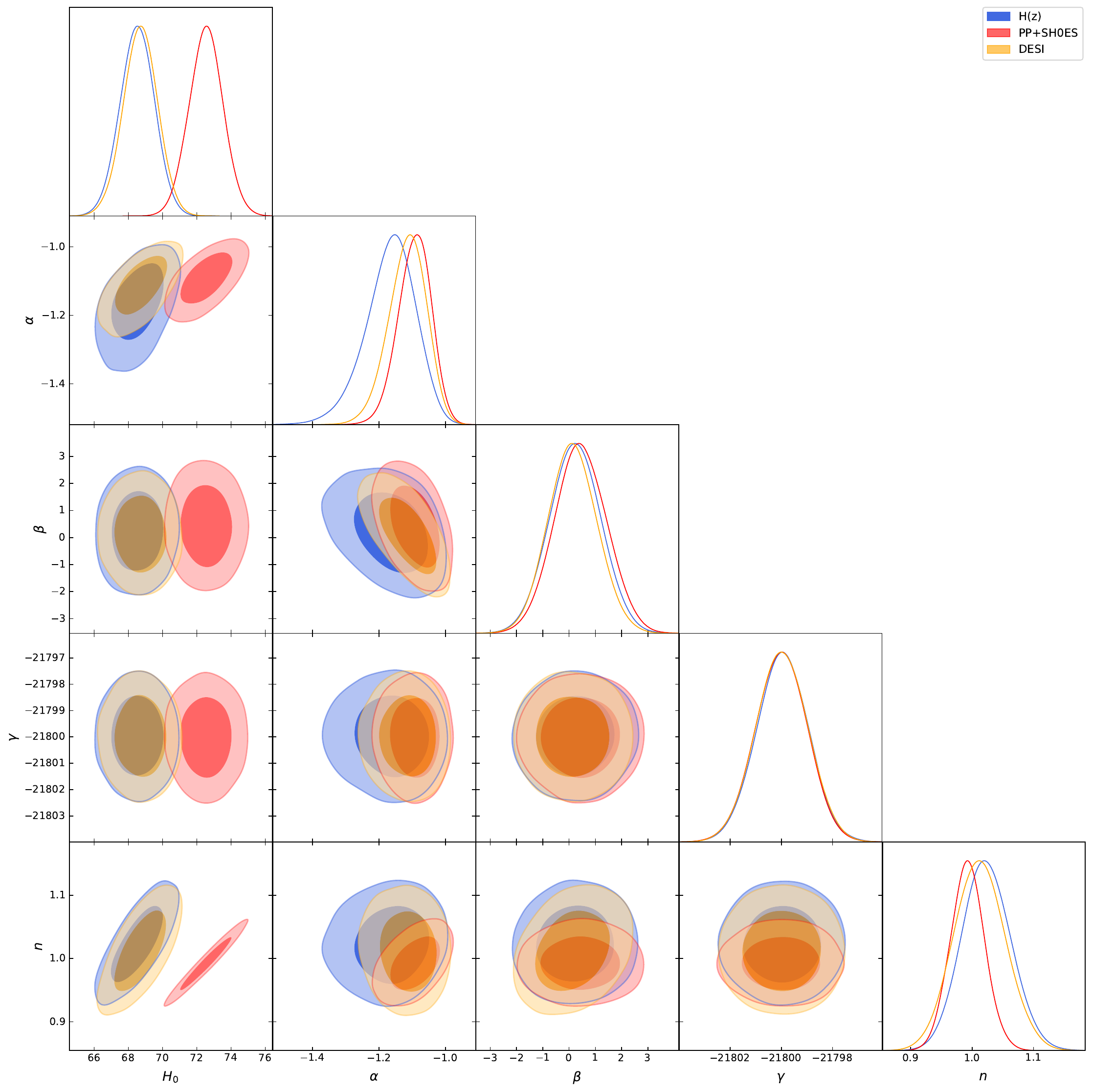}
    \caption{The 1-$\sigma$ and 2-$\sigma$ confidence contours for the $f(Q,T)$ model parameters $(H_0, \alpha, \beta, \gamma, n)$ using Hubble $H(z)$, Pantheon+SH0ES, and DESI data.}
    \label{F_Cont}
\end{figure}

\begin{table}[H]
\centering
\begin{tabular}{lccccccc}
\hline\hline
Dataset & $H_{0}$ [km s$^{-1}$ Mpc$^{-1}$] & $\alpha$ & $\beta$ & $\gamma$ & $n$ & $\chi^2_{\min}$ & $\chi^2_{\text{red}}$ \\
\hline
Priors & $(60,80)$ & $(-10,10)$ & $(-10,10)$ & $(-3,3)\times 10^{4}$ & $(-10,10)$ & $-$ & $-$ \\
$H(z)$ & $68.50 \pm 0.64$ & $-1.166^{+0.082}_{-0.064}$ & $0.24 \pm 0.98$ & $-21800.0 \pm 1.0$ & $1.023 \pm 0.040$ & $21.93$ & $0.84$ \\
Pantheon+SH0ES & $72.54^{+0.26}_{-0.21}$ & $-1.093^{+0.053}_{-0.045}$ & $0.44 \pm 0.99$ & $-21800.0 \pm 1.0$ & $0.993 \pm 0.028$ & $1759.78$ & $1.04$ \\
DESI & $68.7 \pm 0.70$ & $-1.115^{+0.062}_{-0.051}$ & $0.14 \pm 0.93$ & $-21800.0 \pm 1.0$ & $1.012 \pm 0.042$ &  $21.49$ & $3.07$ \\
\hline\hline
\end{tabular}
\caption{Best-fit values of the model parameters $(H_0, \alpha, \beta, \gamma, n)$, minimum chi-square $\chi^2_{\min}$, and reduced chi-square $\chi^2_{\text{red}}$ for the $f(Q,T)$ model with dust matter, obtained using different observational datasets.}
\label{tab1}
\end{table}

\end{widetext}

In the context of the $f(Q,T)$ gravity model, the reduced chi-square ($\chi^2_{\rm red}$) serves as a quantitative measure of the goodness of fit between the theoretical predictions and observational data \cite{Andrae2010}. It is defined as $\chi^2_{\rm red}=\frac{\chi^2_{\rm min}}{\nu}$, where $\chi^2_{\rm min}$ is the chi-square statistic and $\nu=N-n_{\rm par}$ is the number of degrees of freedom (the number of data points minus the number of fitted parameters). A value of $\chi^2_{\rm red} \approx 1$ indicates that the model describes the data well and that the uncertainties are properly estimated. For the $f(Q,T)$ gravity model, the goodness-of-fit analysis reveals the following. For the Hubble parameter data $H(z)$, the reduced chi-square is $\chi^2_{\rm red} \approx 0.84 < 1$, indicating an excellent fit, though the uncertainties may be slightly overestimated. For the combined Pantheon+SH0ES dataset, we obtain $\chi^2_{\rm red} \approx 1.04$, reflecting a very good fit with observational errors consistent with the model predictions. In contrast, the DESI data yield $\chi^2_{\rm red} \approx 3.07$, suggesting some tension or potentially underestimated observational uncertainties.

\section{Cosmographic parameters}
\label{sec5}

The cosmography offers a model-independent way to describe the expansion history of the universe by expanding the scale factor around the present time \cite{Visser_2004}. In a homogeneous and isotropic FLRW universe, the scale factor can be written as a Taylor series:
\begin{equation}
a(t) = \sum_{i=0}^{\infty} \frac{a^{(i)}(t_0)}{i!} \,(t - t_0)^i,    
\end{equation}
where $a^{(i)}(t_0)$ denotes the $i$-th derivative of $a(t)$ evaluated at $t_0$. This expansion naturally defines a set of cosmokinematic parameters \cite{Pan_2018}:
\begin{equation}
H \equiv \frac{\dot a}{a}, \quad
q \equiv -\frac{\ddot a}{aH^2}, \quad
j \equiv \frac{a^{(3)}}{aH^3}, \quad
s \equiv \frac{a^{(4)}}{aH^4}, \quad
l \equiv \frac{a^{(5)}}{aH^5},
\end{equation}
which respectively characterize the Hubble, deceleration, jerk, snap, and lerk of the cosmic expansion. In these definitions, the Hubble parameter $H$ quantifies the rate of cosmic expansion, with its explicit expression already obtained for $f(Q,T)$ gravity (see Eq. (\ref{Hz})). The negative sign in the deceleration parameter is a conventional choice, historically adopted so that a positive value $q>0$ corresponds to a decelerating universe.

\subsection{Deceleration}
The deceleration parameter measures the acceleration or deceleration of the universe. A positive $q>0$ indicates deceleration, while a negative $q<0$ signals cosmic acceleration. It provides the first-order correction to the Hubble expansion and is crucial for distinguishing accelerating models from standard matter-dominated expansion. When expressed in terms of redshift, it takes the form
\begin{equation}
q(z) = (1+z) \frac{1}{H} \frac{dH}{dz} - 1.
\end{equation}

Fig. \ref{F_q} shows the evolution of the deceleration parameter $q(z)$ in the framework of the $f(Q,T)$ gravity model with dust matter. The deceleration parameter $q(z)$ illustrates the transition of the Universe from an early decelerating phase to the current accelerated expansion. At high redshift, $q(z)$ remains positive, indicating decelerated expansion dominated by matter, while at low redshift, it becomes negative, signaling the onset of cosmic acceleration. The present values are $q_0 \sim -0.27$ (Hubble), $q_0 \sim -0.53$ (Pantheon+SH0ES), and $q_0 \sim -0.40$ (DESI), all clearly negative, consistent with the accelerated Universe we observe today \cite{Jesus/2020,Garza/2019,Mamon/2017,Mamon/2018,Luongo_2013}. These results align closely with the $\Lambda$CDM framework. In the far future ($z \to -1$), $q$ tends to $-1$, signaling an asymptotic de Sitter phase. 

\subsection{Jerk}
The jerk parameter captures the rate of change of the acceleration. It is especially useful for identifying departures from the $\Lambda$CDM model, for which $j=1$ at all times. Its expression in terms of redshift is
\begin{equation}
j(z) = (1+z) \frac{dq}{dz} + q(1+2q).
\end{equation}

Fig. \ref{F_j} displays the redshift evolution of the jerk parameter $j(z)$ for the $f(Q,T)$ gravity model with dust matter, using three different datasets ($H(z)$, Pantheon+SH0ES, and DESI). The $j(z)$ jerk parameter contains information about deviations from the $\Lambda$CDM model, where $j=1$ exactly. In the $f(Q,T)$ model, the curves of $j(z)$ remain close to unity throughout cosmic history, implying a $\Lambda$CDM-like behavior but allowing for small deviations at higher redshifts. The present best-fit values are $j_0 \sim 0.97 $ (Hubble), $j_0 \sim 0.99$ (Pantheon+SH0ES), and $j_0 \sim 0.99$ (DESI) \cite{Mamon/2017,Mamon/2018}. In the far future ($z \to -1$), the jerk parameter approaches unity, suggesting that the universe asymptotically tends toward a de Sitter phase consistent with $\Lambda$CDM.

\subsection{Snap}
The snap quantifies the second derivative of the acceleration, providing higher-order information about the cosmic expansion. It is sensitive to subtle deviations from standard cosmological models and can help discriminate between alternative dark energy scenarios. The general expression of this quantity as a function of redshift is
\begin{equation}
s(z) = -(1+z) \frac{dj}{dz} - j(2+3q).
\end{equation}

In Fig. \ref{F_s}, the snap parameter $s$ is displayed as a function of redshift $z$ for the $f(Q,T)$ gravity model with dust matter. The snap parameter $s(z)$ captures higher-order dynamical effects and highlights possible departures from standard cosmology. For large redshifts, $s(z)$ takes negative values, indicating an accelerated expansion phase. Near the present epoch, $s(z)$ exhibits small oscillations around negative values. The current values are $s_0 \sim -1.05$ (Hubble), $s_0 \sim -0.38$ (Pantheon+SH0ES), and $s_0 \sim -0.74$ (DESI) \cite{Jesus_2024}. These negative values support a consistently accelerating Universe in the context of the $f(Q,T)$ model. Interestingly, in the far future ($z \to -1$), the snap parameter evolves to unity, confirming the de Sitter phase of the universe at late times.

\begin{widetext}

\begin{figure}[H]\centering
   \begin{minipage}{0.48\textwidth}
     \includegraphics[width=0.9\linewidth]{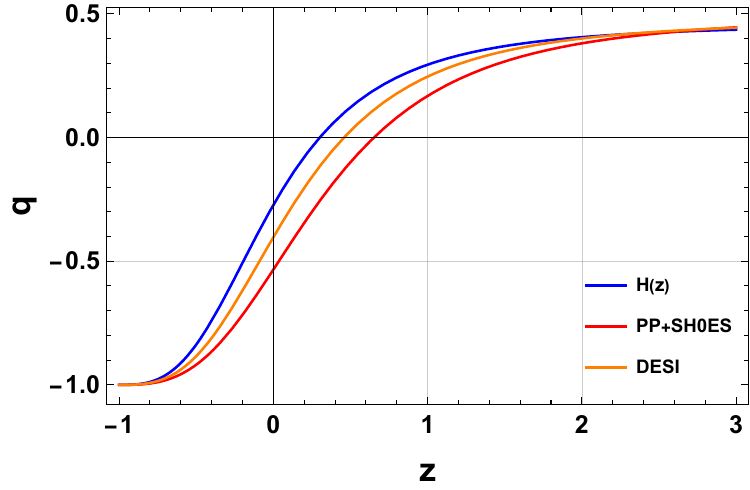}
     \caption{Deceleration parameter $q(z)$ for the $f(Q,T)$ model using Hubble $H(z)$, Pantheon+SH0ES, and DESI data.}\label{F_q}
   \end{minipage} \hfill
   \begin {minipage}{0.48\textwidth}
     \includegraphics[width=0.9\linewidth]{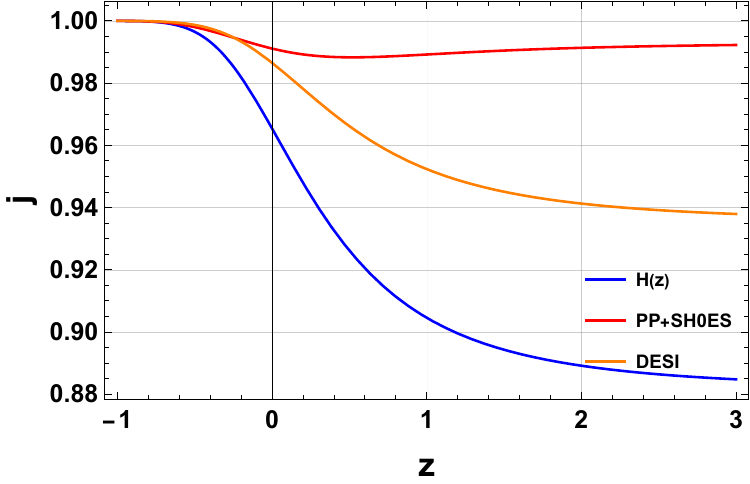}
     \caption{Jerk parameter $j(z)$ for the $f(Q,T)$ model using Hubble $H(z)$, Pantheon+SH0ES, and DESI data.}\label{F_j}
   \end{minipage}
\end{figure}

\begin{figure}[H]\centering
   \begin{minipage}{0.48\textwidth}
     \includegraphics[width=0.9\linewidth]{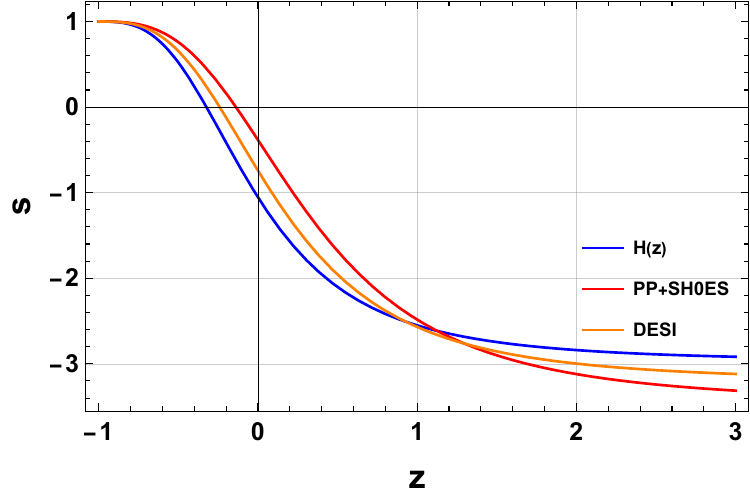}
     \caption{Snap parameter $s(z)$ for the $f(Q,T)$ model using Hubble $H(z)$, Pantheon+SH0ES, and DESI data.}\label{F_s}
   \end{minipage} \hfill
   \begin {minipage}{0.48\textwidth}
     \includegraphics[width=0.9\linewidth]{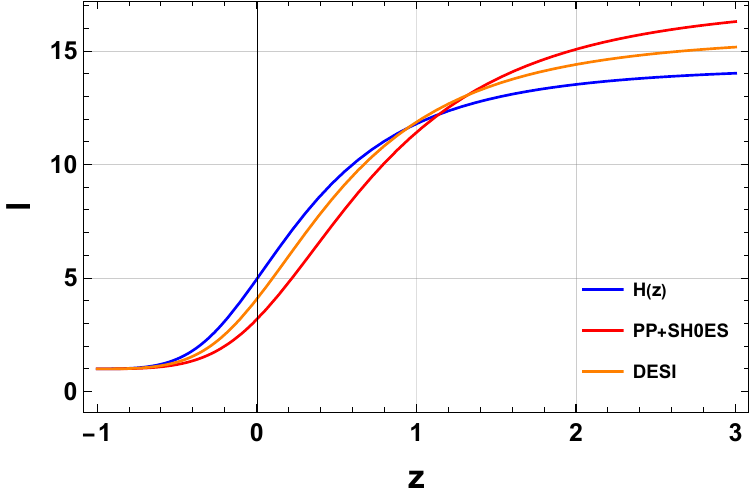}
     \caption{Lerk parameter $l(z)$ for the $f(Q,T)$ model using Hubble $H(z)$, Pantheon+SH0ES, and DESI data.}\label{F_l}
   \end{minipage}
\end{figure}

\begin{table}[H]
\centering
\begin{tabular}{lcccc}
\hline\hline
Dataset & $q_{0}$ & $j_{0}$ & $s_{0}$ & $l_{0}$ \\
\hline
$H(z)$ & $-0.27 \pm 0.35$ & $0.97 \pm 0.08$ & $-1.05 \pm 0.82$ & $5.05 \pm 2.0$ \\
Pantheon+SH0ES & $-0.53 \pm 0.36$ & $0.99 \pm 0.03$ & $-0.38 \pm 1.01$ & $3.18 \pm 1.03$ \\
DESI & $-0.40 \pm 0.46$ & $0.99 \pm 0.06$ & $-0.74 \pm 1.19$ & $4.04 \pm 2.21$ \\
\hline\hline
\end{tabular}
\caption{Present values of the cosmographic parameters derived from the best-fit nonlinear $f(Q,T)$ model parameters using different observational datasets.}
\label{tab2}
\end{table}

\end{widetext}

\subsection{Lerk}
The lerk parameter (sometimes called crackle) captures the third derivative of the acceleration. Although harder to measure observationally, it offers even finer knowledge about the expansion history, refining cosmographic constraints on complex cosmological models. In terms of redshift, it can be generally written as
\begin{equation}
l(z) = -(1+z) \frac{ds}{dz} - s(3+4q).
\end{equation}

In Fig. \ref{F_l}, the behavior of the lerk parameter $l$ is shown as a function of the redshift $z$ for the $f(Q,T)$ gravity model with dust matter. The lerk parameter $l(z)$ encodes even higher-order corrections to the cosmic expansion. In the analyzed model, $l(z)$ remains strictly positive over the entire redshift range, which is generally associated with a smooth and stable accelerated expansion. The present values are $l_0 \sim 5.05$ (Hubble), $l_0 \sim 3.18$ (Pantheon+SH0ES), and $l_0 \sim 4.04$ (DESI). A positive $l_0$ further confirms that the $f(Q,T)$ scenario with dust matter supports the current acceleration of the Universe.

\section{Conclusion} \label{sec6}

The accelerating expansion of the Universe remains one of the most intriguing challenges in modern cosmology. Over the past decades, a variety of dynamical dark energy models and modified theories of gravity have been proposed to account for this phenomenon, each offering a different perspective on the underlying physics. Despite these efforts, a fully satisfactory and universally accepted description of cosmic acceleration has yet to be achieved, motivating ongoing research into alternative frameworks. In this work, we have investigated the nonlinear $f(Q,T)$ gravity model, $f(Q,T) = \alpha Q^n + \beta T + \gamma$, which represents the nontrivial coupling between geometry and matter \cite{Xu/2019}. We focus on a dust-dominated Universe and constrain the model parameters $(H_0, \alpha, \beta, \gamma, n)$ using the latest cosmological observations, including Hubble parameter measurements $H(z)$, the Pantheon+SH0ES dataset, and DESI BAO data.

The analysis (Fig. \ref{F_Cont} and Tab. \ref{tab1}) shows that the Hubble constant $H_0$ varies across datasets, reflecting the well-known Hubble tension: $H_0 \sim 68.50$ (Hubble), $H_0 \sim 72.54$ (Pantheon+SH0ES), and $H_0 \sim 68.7$ (DESI). The parameter $\alpha$ is tightly constrained around $-1$, while $\beta$ remains weakly constrained, and the offset $\gamma \simeq -2.18 \times 10^4$ is robustly determined. The nonlinear exponent $n$ introduces an additional degree of freedom that enhances the model’s flexibility in reproducing observational trends. Further, the reduced chi-square values indicate a perfect fit for the Hubble and Pantheon+SH0ES data, with some tension for the DESI dataset, likely due to underestimated uncertainties. The cosmographic analysis of the deceleration, jerk, snap, and lerk parameters (Figs. \ref{F_q}-\ref{F_l} and Tab. \ref{tab2}) further confirms that the model consistently describes an accelerating Universe. The present values, $q_0$ from $-0.27$ to $-0.53$, $j_0$ from $0.97$ to $0.99$, $s_0$ from $-0.38$ to $-1.05$, and $l_0$ from $3.18$ to $5.055$, align closely with $\Lambda$CDM predictions, while the higher-order parameters suggest minor deviations at earlier times. Notably, in the far future ($z \to -1$), all cosmographic parameters approach values consistent with a de Sitter phase, confirming that the model predicts a stable accelerated expansion.

Hence, the nonlinear $f(Q,T)$ model with dust matter provides a simple yet effective framework to describe late-time cosmic acceleration, accommodate the Hubble tension, and reproduce $\Lambda$CDM-like behavior with subtle deviations, demonstrating its viability as an alternative gravitational scenario for cosmology.

\section*{Acknowledgements}
This research was funded by the Science Committee of the Ministry of Science and Higher Education of the Republic of Kazakhstan (Grant No. AP32720463).

\section*{Data Availability Statement}
There are no new data associated with this article.


\begin{thebibliography}{90}

\bibitem{Perlmutter/1999} S. Perlmutter et al., \textit{Astrophys. J.} \textbf{517}, 565 (1999).

\bibitem{Riess/1998} A.G. Riess et al., \textit{Astron. J.} \textbf{116}, 1009 (1998).  

\bibitem{Riess/2004} A.G. Riess et al., \textit{Astrophys. J.} \textbf{607}, 665 (2004).  

\bibitem{Spergel/2003} D.N. Spergel et al., \textit{Astrophys. J. Suppl.} \textbf{148}, 175 (2003).  

\bibitem{Koivisto/2006} T. Koivisto, D.F. Mota, \textit{Phys. Rev. D} \textbf{73}, 083502 (2006).  

\bibitem{Daniel/2008} S.F. Daniel et al., \textit{Phys. Rev. D} \textbf{77}, 103513 (2008).  

\bibitem{Weinberg/1989} S. Weinberg, \textit{Rev. Mod. Phys.} \textbf{61}, 1 (1989).

\bibitem{Hinshaw/2013} G. Hinshaw et al., \textit{Astrophys. J. Suppl.} \textbf{208}, 19 (2013).  

\bibitem{Zhao/2012} G.-B. Zhao et al., \textit{Phys. Rev. Lett.} \textbf{109}, 171301 (2012).  

\bibitem{Ratra/1988} B. Ratra, P.E.J. Peebles, \textit{Phys. Rev. D} \textbf{37}, 3406 (1988).  

\bibitem{Caldwell/1998} R. Caldwell, R. Dave, P.J. Steinhardt, \textit{Phys. Rev. Lett.} \textbf{80}, 1582 (1998).  

\bibitem{Caldwell/2002} R.R. Caldwell, \textit{Phys. Lett. B} \textbf{545}, 23 (2002).  

\bibitem{Caldwell/2003} R.R. Caldwell, M. Kamionkowski, N.N. Weinberg, \textit{Phys. Rev. Lett.} \textbf{91}, 071301 (2003).  

\bibitem{Feng/2005} B. Feng, X. Wang, X. Zhang, \textit{Phys. Lett. B} \textbf{607}, 35 (2005).  

\bibitem{Huang/2005} G.Q. Huang, M. Li, \textit{J. Cosmol. Astropart. Phys.} \textbf{03}, 001 (2005).  
\bibitem{Afshordi/2007} N. Afshordi, D.J.H. Chung, G. Geshnizjani, \textit{Phys. Rev. D} \textbf{75}, 083513 (2007).  

\bibitem{Buchdahl/1970} H.A. Buchdahl, \textit{Mon. Not. R. Astron. Soc.} \textbf{150}, 1 (1970).

\bibitem{Barrow/1983} J.D. Barrow and A.C. Ottewill, \textit{J. Phys. A: Math. Gen.} \textbf{16}, 2757 (1983).

\bibitem{Boehmer/2008} C.G. Boehmer, T. Harko, and F.S.N. Lobo, \textit{Astropart. Phys.} \textbf{29}, 386 (2008).

\bibitem{Moller/1961} C. Möller, \textit{Mat. Fys. Skr. Dan. Vid. Selsk.} \textbf{1}, 10 (1961).

\bibitem{Pellegrini/1963} C. Pellegrini and J. Plebanski, \textit{Mat. Fys. Skr. Dan. Vid. Selsk.} \textbf{2}, 4 (1963).

\bibitem{Hayashi/1979} K. Hayashi and T. Shirafuji, \textit{Phys. Rev. D} \textbf{19}, 3524 (1979).

\bibitem{Linder/2010} E.V. Linder, \textit{Phys. Rev. D} \textbf{81}, 127301 (2010).

\bibitem{Boehmer/2012} C.G. Boehmer, T. Harko, and F.S.N. Lobo, \textit{Phys. Rev. D} \textbf{85}, 044033 (2012).

\bibitem{Harko/2014} T. Harko, F.S.N. Lobo, G. Otalora, and E.N. Saridakis, \textit{Phys. Rev. D} \textbf{89}, 124036 (2014).

\bibitem{Nester/1999} J.M. Nester and H.-J. Yo, \textit{Chin. J. Phys.} \textbf{37}, 113 (1999), arXiv:gr-qc/9809049.  

\bibitem{Adak/2006} M. Adak, M. Kalay, and O. Sert, \textit{Int. J. Mod. Phys. D} \textbf{15}, 619 (2006).  

\bibitem{Jimenez/2018} J.B. Jimenez, L. Heisenberg, and T. Koivisto, \textit{Phys. Rev. D} \textbf{98}, 044048 (2018).

\bibitem{Jimenez/2020} J.B. Jimenez, L. Heisenberg, T. Koivisto, and S. Pekar, \textit{Phys. Rev. D} \textbf{101}, 103507 (2020).  
\bibitem{Koussour/2023a} M. Koussour, S. Arora, D.J. Gogoi, M. Bennai, and P.K. Sahoo, \textit{Nucl. Phys. B} \textbf{990}, 116158 (2023).  

\bibitem{Koussour/2023b} M. Koussour, S.K.J. Pacif, M. Bennai, and P.K. Sahoo, \textit{Fortschr. Phys.} \textbf{71}, 2200172 (2023).  

\bibitem{Koussour/2022a} M. Koussour, K. El Bourakadi, S.H. Shekh, S.K.J. Pacif, and M. Bennai, \textit{Ann. Phys.} \textbf{445}, 169092 (2022).  

\bibitem{Koussour/2022b} M. Koussour, S.H. Shekh, and M. Bennai, \textit{J. High Energy Astrophys.} \textbf{35}, 43 (2022).  

\bibitem{Koussour/2022c} M. Koussour, S.H. Shekh, and M. Bennai, \textit{Phys. Dark Univ.} \textbf{36}, 101051 (2022).  

\bibitem{Koussour/2024a} M. Koussour, S.H. Shekh, M. Bennai, and N. Myrzakulov, \textit{Mod. Phys. Lett. A} \textbf{39}, 2450023 (2024).  

\bibitem{Koussour/2024b} M. Koussour and N. Myrzakulov, \textit{Eur. Phys. J. Plus} \textbf{139}, 799 (2024).  

\bibitem{Xu/2019} Y. Xu et al., \textit{Eur. Phys. J. C} \textbf{79}, 708 (2019).  

\bibitem{Harko/2011} T. Harko et al., \textit{Phys. Rev. D} \textbf{84}, 024020 (2011).  

\bibitem{Arora/2020} S. Arora, S.K.J. Pacif, S. Bhattacharjee, and P.K. Sahoo, \textit{Phys. Dark Univ.} \textbf{30}, 100664 (2020).  

\bibitem{Najera/2022} A. Nájera and A. Fajardo, \textit{JCAP} \textbf{03}, 020 (2022).  

\bibitem{Shiravand/2022} M. Shiravand, S. Fakhry, and M. Farhoudi, \textit{Phys. Dark Univ.} \textbf{37}, 101106 (2022).  

\bibitem{Koussour/2024} M. Koussour, A. Errehymy, O. Donmez, K. Myrzakulov, M.A. Khan, B. Çil, and E. Güdekli, \textit{Phys. Dark Univ.} \textbf{45}, 101527 (2024).  

\bibitem{ElBourakadi/2023} K. El Bourakadi, M. Koussour, G. Otalora, M. Bennai, and T. Ouali, \textit{Phys. Dark Univ.} \textbf{41}, 101246 (2023).  

\bibitem{Koussour/2022} M. Koussour, N. Myrzakulov, S.H. Shekh, and M. Bennai, \textit{Int. J. Mod. Phys. D} \textbf{31}, 2250115 (2022).  

\bibitem{Loo/2023} T.H. Loo, M. Koussour, and A. De, \textit{Ann. Phys.} \textbf{454}, 169333 (2023).  

\bibitem{Agrawal/2021} A.S. Agrawal, L. Pati, S.K. Tripathy, and B. Mishra, \textit{Phys. Dark Univ.} \textbf{33}, 100863 (2021).  

\bibitem{Xu/2020} Y. Xu, T. Harko, S. Shahidi, and S.D. Liang, \textit{Eur. Phys. J. C} \textbf{80}, 449 (2020).

\bibitem{Yang/2021} J.Z. Yang, S. Shahidi, T. Harko, and S.D. Liang, \textit{Eur. Phys. J. C} \textbf{81}, 111 (2021).

\bibitem{Loo/2023} T.H. Loo, R. Solanki, A. De, and P.K. Sahoo, \textit{Eur. Phys. J. C} \textbf{83}, 261 (2023).

\bibitem{Weinberg/2013} S. Weinberg, \textit{Gravitation and cosmology: principles and applications of the general theory of relativity} (John Wiley \& Sons, 2013).

\bibitem{Dodelson/2024} S. Dodelson and F. Schmidt, \textit{Modern Cosmology} (Elsevier, 2024).

\bibitem{emcee} D. F. Mackey et al.,  Publ. Astron. Soc. Pac., \textbf{125}, 306 (2013).

\bibitem{Moresco/2012} M. Moresco et al., \textit{J. Cosmol. Astropart. Phys.} \textbf{08}, 006 (2012). 

\bibitem{Moresco/2015} M. Moresco, \textit{Mon. Not. R. Astron. Soc.} \textbf{450}, L16-L20 (2015).

\bibitem{Moresco/2016} M. Moresco, \textit{J. Cosmol. Astropart. Phys.} \textbf{05}, 014 (2016).

\bibitem{Moresco/2020} M. Moresco et al., \textit{Astrophys. J.} \textbf{898}, 82 (2020).
\bibitem{Scolnic_2022} D. M. Scolnic et al., \textit{Astrophys. J.} \textbf{938}, 113 (2022).

\bibitem{Brout_2022} D. Brout et al., {\it ApJ} {\bf 938}, 110 (2022).

\bibitem{BAO1} DESI collaboration, \textit{arXiv:2404.03000} (2024).

\bibitem{BAO2} A.G. Adame et al., \textit{J. Cosmol. Astropart. Phys.} \textbf{01}, 124 (2025).

\bibitem{BAO3} A.G. Adame et al., \textit{J. Cosmol. Astropart. Phys.} \textbf{02}, 021 (2025).

\bibitem{D.J.} D.J. Eisenstein et al.. \textit{Astrophys. J.} \textbf{633}, 560 (2005).


\bibitem{Planck_2018} N. Aghanim et al. (Planck Collaboration), {\it A\&A} {\bf 641}, A6 (2020).

\bibitem{Riess_2016} A. G. Riess et al., {\it ApJ} {\bf 826}, 56 (2016).

\bibitem{Riess_2022} A. G. Riess et al., {\it ApJL} {\bf 934}, L7 (2022).

\bibitem{Yang_2019} W. Yang et al., {\it Mon. Not. R. Astron. Soc.} {\bf 490}, 2071 (2019).

\bibitem{Valentino_2020} E. Di Valentino, A. Melchiorri, and J. Silk, {\it JCAP} {\bf 2020}(01), 013 (2020).

\bibitem{Mukherjee_2021} P. Mukherjee and N. Banerjee, {\it Eur. Phys. J. C} {\bf 81}, 36 (2021).

\bibitem{Campo_2012} S. del Campo, I. Duran, R. Herrera, and D. Pavon, {\it Phys. Rev. D} {\bf 86}, 083509 (2012).

\bibitem{Cao_2021} S. Cao, J. Ryan, and B. Ratra, {\it Mon. Not. R. Astron. Soc.} {\bf 504}, 300 (2021).

\bibitem{Andrae2010} R. Andrae, T. Schulze-Hartung, and P. Melchior, \textit{arXiv:1012.3754} (2010).

\bibitem{Visser_2004} M. Visser, {\it Class. Quantum Grav.} {\bf 21}, 2603 (2004).

\bibitem{Pan_2018} S. Pan, A. Mukherjee, and N. Banerjee, {\it Mon. Not. R. Astron. Soc.} {\bf 477}, 1189 (2018).

\bibitem{Jesus/2020} J.F. Jesus, et al., \textit{J. Cosmol. Astropart. Phys.} \textbf{053}, 04 (2020).

\bibitem{Garza/2019} J. R. Garza, et al., \textit{Eur. Phys. J. C} \textbf{79}, 890 (2019).

\bibitem{Mamon/2017} A. Al Mamon, and S. Das, \textit{Eur. Phys. J. C} \textbf{77}, 495 (2017).

\bibitem{Mamon/2018} A. Al Mamon, and K. Bamba, \textit{Eur. Phys. J. C} \textbf{78}, 862 (2018).

\bibitem{Luongo_2013} O. Luongo, \textit{Mod. Phys. Lett. A} \textbf{29}, 1350080 (2013).

\bibitem{Jesus_2024} J.F. Jesus, D. Benndorf, A.A. Escobal, and S.H. Pereira, \textit{Mon. Not. R. Astron. Soc.} \textbf{528}, 1573 (2024).



\end{thebibliography}
\end{document}